\documentclass[aps,amsmath,preprint,prd,showpacs,nofootinbib]{revtex4}
\newcommand*{\ket}[1]{\ensuremath{|#1\rangle}}
\usepackage{amssymb,axodraw,graphics,graphicx,color,dcolumn,bm,slashed}

\def\roughly#1{\mathrel{\raise.3ex\hbox
{$#1$\kern-.75em\lower1ex\hbox{$\sim$}}}}

\begin{document}

\title{\boldmath Peccei-Quinn symmetry as the origin of Dirac Neutrino Masses}
\author{Chian-Shu~Chen$^{1,3}$\footnote{chianshu@phys.sinica.edu.tw} and Lu-Hsing Tsai$^{2}$\footnote{lhtsai@phys.nthu.edu.tw}}

\affiliation{$^{1}$Physics Division, National Center for Theoretical Sciences, Hsinchu, Taiwan 300\\ 
$^{2}$Department of Physics, National Tsing Hua University, Hsinchu, Taiwan 300\\
$^{3}$Institute of Physics, Academia Sinica, Taipei, Taiwan 115}

\begin{abstract}
We propose a model of Dirac neutrino masses generated at one-loop level. The origin of this mass is induced
from Peccei-Quinn symmetry breaking which was proposed to solve the so-called strong CP problem in QCD,
therefore, the neutrino mass is connected with the QCD scale, $\Lambda_{\rm QCD}$. We also study the
parameter space of this model confronting with neutrino oscillation data and leptonic rare decays. The
phenomenological implications to leptonic flavor physics such as the electromagnetic moment of charged leptons
and neutrinos are studied. Axion as the dark matter candidate is one of the byproduct in our scenario.
Di-photon and Z-photon decay channels in the LHC Higgs search are investigated. We show that the effects
of singly charged singlet scalar can be distinguished from the general two Higgs doublet model.
\end{abstract}
\pacs{14.60.Pq, 12.60.-i, 14.80.-j, 14.80.Mz}
%\keywords{}

\maketitle

\section{Introduction}
\label{sec:intro}
Small quantities arising in physics usually requires the use of new symmetries for explanations~\cite{t'hooft}.
A good example of these is the Peccei-Quinn (PQ) symmetry that plays the role for the solution of the strong
CP-problem~\cite{Peccei:1977hh,Peccei:1977ur}, in which a $\theta$-angle appears in the QCD lagrangian,
%\begin{eqnarray}
$\mathcal{L}_{QCD} \in \theta\frac{g^2_{S}}{32\pi^2}G^{a}_{\mu\nu}\tilde{G}^{a\mu\nu}$, and CP is
violated\footnote{The $G^{a}_{\mu\nu}$ are the $SU(3)_C$
gauge fields with $a = 1,2,...,8$ and $\tilde{G}^{a\mu\nu} = \frac{1}{2}\epsilon^{\mu\nu\alpha\beta}G^{a}_{\alpha\beta}$.}~\cite{'tHooft:1976up,Jackiw:1976pf,Callan:1976je}. The instanton solution to the gluon field equations
satisfies $n = \frac{1}{32\pi^2}\int d^4xG^{a}_{\mu\nu}\tilde{G}^{a\mu\nu}$ with $n$'s being integers and representing
topological charges~\cite{Belavin:1975fg}. The QCD vacuum state hence can be parametrized as
$\ket{\theta} = \sum_{n = -\infty}^{n = \infty}e^{in\theta}\ket{n}$ where $\theta$ is periodic with period $2\pi$.
Furthermore for nonzero quark masses the chiral anomaly relates the weak phase in quark masses to the
QCD $\theta$-term. One can parametrize the $\theta$ angle
as $\bar{\theta} = \theta - {arg}\{{\rm{det}}m_{q}\}$~\cite{Adler:1969gk,Bell:1969ts}. It induces a
neutron electric dipole moment (EDM)~\cite{Baluni:1978rf,Crewther:1979pi,Cea:1984qv,Schnitzer:1983pb,Musakhanov:1984qy}
and the current experimental upper bound set the best constraint on $\bar{\theta}$ to be smaller
than $0.6\times10^{-10}$~\cite{Baker:2006ts}. This extremely suppressed quantity is called the strong CP problem.
The Peccei-Quinn solution to the strong CP problem postulates a global chiral $U(1)_{PQ}$ symmetry and
makes $\bar{\theta}$ a dynamical variable, and the shift symmetry of the Nambu-Goldstone boson, axion,
corresponding to $U(1)_{PQ}$~\cite{Weinberg:1977ma,Wilczek:1977pj} will set $\bar{\theta}$ zero
at classical potential~\cite{Vafa:1984xg}. At one-loop level the chiral anomaly will break the shift symmetry.
As a result the axion is not massless but requires a small mass
$m_{a} \simeq \frac{\sqrt{m_{u}m_{d}}}{(m_{u} + m_{d})}\frac{f_{\pi}m_{\pi}}{f_{a}} \sim \frac{\Lambda^2_{QCD}}{f_{a}}$~\cite{Peccei:2006as,Kim:1986ax,Cheng:1987gp}. Here $f_{a}$ is the $U(1)_{PQ}$ breaking scale and
$f_{\pi}$ is the pion decay constant. The laboratory~\cite{pdg} and outer space~\cite{Andriamonje:2007ew,Arik:2011rx,Inoue:2008zp,Asztalos:2006kz,Asztalos:2009yp} searches have set $10^{9}~{\rm GeV}\lesssim f_{a} \lesssim 3\times10^{11}~{\rm{GeV}}$
as the allowed regions, therefore, the axion window is $3\times10^{-3}~{\rm eV} > m_{a} > 10^{-6}~{\rm eV}$.

On the other hand, another small quantity that puzzles high energy physicists is the masses of neutrinos measured from
neutrino oscillation experiments~\cite{pdg}. The key point to understand neutrino physics lies on whether the neutrinos are
Dirac fermions or Majorana fermions. This ambiguity comes from the fact of zero electric charge carried by neutrinos.
The tiny neutrino masses may be explained in terms of lepton number ($L$) symmetry which is a global $U(1)$ quantum
number tagged on lepton sectors in the standard model (SM). If $U(1)_{L}$ is broken one can write the dimension-5 Weinberg
operator to generate neutrino masses $m_{\nu} \propto \frac{HHLL}{\Lambda_{L}}$~\cite{Weinberg:1979sa}, where $H$ and $L$
are the SM Higgs and the left-handed lepton fields respectively, and $\Lambda_{L}$ is the breaking scale of $U(1)_{L}$. In this case
neutrinos are regarded as Majorana fermions. However, the Majorana or Dirac nature of the neutrinos is unknown and is awaiting
for the experimental determination from some lepton number violating processes such as the neutrinoless double beta decay.
It is important to consider the possibility that Dirac neutrino masses may also connect with some global symmetry and how
those small quantities we observe in physics are related to each other.\footnote{The interesting models of connecting Dirac neutrino mass with leptogenesis are studied, for example, in Refs.~\cite{Gu:2006dc,Gu:2007ug,Gu:2012fg}.} In this paper we propose a simple Dirac neutrino mass
model which is generated by PQ symmetry breaking, and hence the neutrino masses are closely related with axion mass.\footnote{In Ref.~\cite{Davidson:1987tr}, the PQ symmetry and Dirac neutrino masses are connected in the so-called universal seesaw model. Similar idea of linking Majorana neutrino masses with PQ symmetry was also studied in Ref.~\cite{Mohapatra:1982tc, Shafi:1984ek, Langacker:1986rj, Shin:1987xc, Geng:1988gr, He:1988dm, Geng:1988nc, Ma:2001ac, Bertolini:1990vz, Arason:1990sg}.}

This paper is organized as follows : in section~\ref{sec:model} we propose the Dirac mass model which is embedded 
with PQ symmetry. Section~\ref{sec:3} we consider leptonic rare decays and neutrino oscillations data to investigate 
the parameter space of the model. Some phenomenological implications to LHC Higgs search, dark matter, and leptonic 
flavor physics of this model are discussed
in section~\ref{sec:4}. Then we conclude our results in section~\ref{sec:5}.
\section{The Model}
\label{sec:model}
Particle content and their quantum numbers of the model are listed in Table~\ref{model}. The $Y$, $L$ and PQ 
represent the hypercharge, lepton number, and $U(1)_{PQ}$ charge respectively\footnote{In this paper two additional 
global symmetries $U(1)_L$ and $U(1)_{PQ}$ are imposed in the model. It turns out the two symmetries are not independent such that $U(1)_L$ can be generated 
accidentally by some particular choice of $U(1)_{PQ}$ charges. One example is to take the PQ charges as 
$H_1:2$, $H_2:-2$, $L_L:1$, $\nu_R:4$, $l_R:-1$, $a:1$, $s_1:-2$, and $s_2:-3$. We found it is a generic feature of assigning the large ratios of the PQ charges for some particles in order to bridge the two global symmetries nontrivially.}. Two Higgs doublets are introduced since the existence of
the PQ-symmetry and one need two independent chiral transformation for the up-type and down-type fermion, that is, one scalar doublet $H_1$ couples to $d_R$ and $l_R$, while the other one $H_2$ only couples to $u_R$ by setting opposite PQ charges to doublet scalars. We consider
the scenario that neutrinos are Dirac fermion, with three right-handed neutrinos $\nu_{R_{i}}~(i=1-3)$
assigned to our model, and hence the theory is lepton number conserved. $s_{1}^{+}$ and $s_{2}^{+}$ are $SU(2)_{L}$-singlet
charged scalars, and $a$ is the axion field which is the Nambu-Goldstone boson of the spontaneously broken $U(1)_{PQ}$.
Notice that $\nu_{R}$'s are complete neutral under gauge symmetries and PQ symmetry and only carry the $L$ quantum number.
The Dirac neutrino mass term is forbidden by the PQ-symmetry at the tree level and is generated at one-loop level after the PQ
symmetry breaking by utilizing the charged scalars $s_{1}^{+}$ and $s^{+}_{2}$.
\begin{table}[tbp]
\centering
\begin{tabular}{|c|cccccccc|}
  \hline
   & $L_L$ & $l_R$ &$H_1$ &$H_2$ &$\nu_{R_i}$&$s_1^+$&$s_2^+$&$a$\\
   \hline
  $Y$ & $-\frac{1}{2}$ & -1 &$\frac{1}{2}$&$\frac{1}{2}$&0&1&1&0\\
  $L$ & 1 & 1&0&0&1&-2&-2&0 \\
  PQ & 0 & -2& 2 & -2 &0&0&2&-2 \\
  \hline
\end{tabular}
\caption{\label{model}Quantum numbers of $U(1)_{PQ}$ and gauge symmetries for leptons and scalars.}
\end{table}
The new Yukawa interactions of the model for leptons are given by
\begin{equation}
\mathcal{L} = y_{\alpha\beta}\overline{(L_{L_{\alpha}})}l_{R_{\beta}}H_1+f_{\alpha\beta}\overline{L^c_{L_{\alpha}}}i\sigma_{2}(L_{L_{\beta}})s_1^+
+h_{\alpha i}\overline{l_{R_{\alpha}}^c}\nu_{R_i}s_2^+ +\mathrm{h.c.}, %+V(H_1,H_2,s_1^+,s_2^+,D)\,.
\end{equation}
where $c$ denotes charged conjugation, $\alpha,\beta = e, \mu, \tau$, and $\sigma_{i}~(i=1-3)$ are the Pauli matrices.
In general, $y$ and $h$ are complex matrices, and $f$ is an antisymmetric matrix due to the Fermi statistics. One can choose
the basis of leptonic mass eigenstates $L_{L}$, $l_{R}$ such that $y$ is a diagonalized matrix. Also $f_{ij}$ can be chosen to
be real by rephasing $L_L$ and by transferring the phases into $l_R$. Therefore only $h_{ij}$ are complex.
The scalar potential can be written as
\begin{equation}
\begin{split}
V &=-\mu_1^2 H_1^\dagger H_1 -\mu_2^2 H_2^\dagger H_2 -\mu_a^2 |a|^2 + \mu_{s1}^2 |s_1|^2+\mu_{s2}^2 |s_2|^2 \\
&+\lambda_1(H_1^\dagger H_1)(H_1^\dagger H_1)+\lambda_2(H_2^\dagger H_2)(H_2^\dagger H_2)
+\lambda_3(H_1^\dagger H_1)(H_2^\dagger H_2)+\lambda_4(H_1^\dagger H_2)(H_2^\dagger H_1) \\
&+(H_1^\dagger H_1)[d_{1}|s_1|^2+d_{2}|s_2|^2+d_{a}|a|^2)]
+(H_2^\dagger H_2)[g_{1}|s_1|^2+g_{2}|s_2|^2+g_{a}|a|^2] \\
&+h_1|s_1|^4+h_2|s_2|^4+h_3|s_1|^2|s_2|^2+h_a|a|^4+|a|^2(h_{a1}|s_1|^2+h_{a2}|s_2|^2) \\
&+[h_{5}(H_2^\dagger H_1)a^2+\mu s_1^- s_2^+a+\mathrm{h.c.}].
\end{split}\label{potential}
\end{equation}
All parameters in the potential are real. Though $h_{5}$ and $\mu$ are in general complex parameters, one
can absorb their phases by redefining $(s_1^\dagger s_2)$, $(H_2^\dagger H_1)$ and $a$ respectively. For the
invisible axion, $d_a$, $g_a$, $h_{a1}$, $h_{a2}$, $h_5$, $\mu$ should be very small to make other scalar mass
not too heavy~\cite{Dine:1981rt,Zhitnitsky:1980tq,Kim:1979if,Shifman:1979if}. The details about the scalar mass spectrum are put in the Appendix.

\begin{figure}[tbp]
  \centering
  \includegraphics[width=85mm,height=60mm]{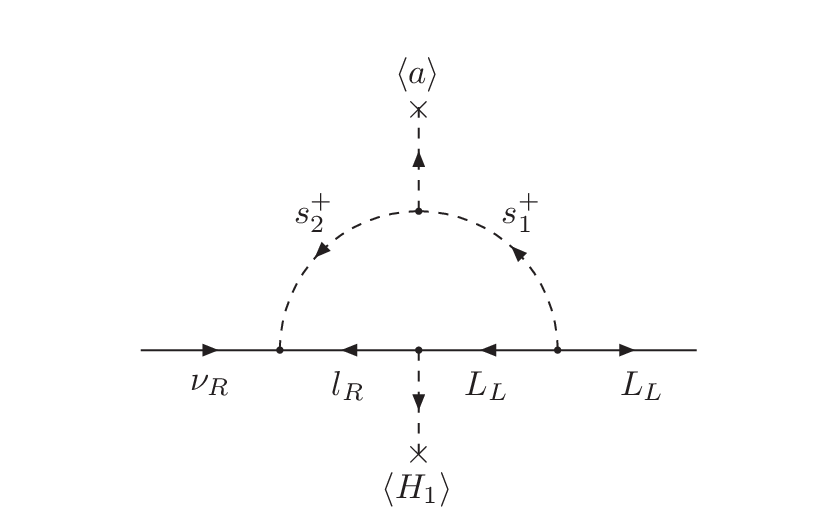}
  \caption{Induced Dirac neutrino mass}\label{Fig_nmass}
\end{figure}

The leading contribution to the Dirac neutrino mass at one-loop level is shown in Fig.~\ref{Fig_nmass} and is given by
\begin{equation}
(M_\nu)_{\alpha i} = -\frac{\mu f_{a}}{8\pi^2m_{\beta}}f_{\alpha\beta}h_{\beta i} I(m_{s1}^2,m_{s2}^2,m_{\beta}^2)\,,
\end{equation}
%{\bf For the $s_1^+$, $s_2^+$ states}
%\begin{eqnarray}
%-i\mathcal{M}&=&\mu^{4-d}\int \frac{d^d k}{(2\pi)^d}\bar{\nu}_i (i2f'_{ik} P_R)\frac{i}{\slashed{k}+\slashed{p}-m_k}(ih_{kj} P_R) \nu_j\frac{i}{k^2-m_{s1}^2}(i \mu_Dv_D)\frac{i}{k^2-m_{s2}^2}\nonumber\\
%&=& \mu^{4-d}\int \frac{d^d k}{(2\pi)^d}\frac{N(k)}{((k+p)^2-m_k^2)(k^2-m_{s1}^2)(k^2-m_{s2}^2)}\nonumber\\
%&=&\mu^{4-d}\int \frac{d^d k}{(2\pi)^d}\frac{N(k)}{m_{s1}^2-m_{s2}^2}
%\bigg[\frac{1}{((k+p)^2-m_k^2)(k^2-m_{s1}^2)}-\frac{1}{((k+p)^2-m_k^2)(k^2-m_{s2}^2)}\bigg]\nonumber\\
%&=&\mu^{4-d}\int dx\int \frac{d^d k}{(2\pi)^d}\frac{N(k-xp)}{m_{s1}^2-m_{s2}^2}
%\bigg[\frac{1}{(k^2-M_1^2)^2}-\frac{1}{(k^2-M_2^2)^2}\bigg]\nonumber\\
%\end{eqnarray}
%where $M_{1,2}^2=xm_k^2+(1-x)m_{s1,2}^2-x(1-x)p^2$, and $N(k)$ is defined by
%\begin{eqnarray}
%N(k)=-\mu_Dv_D\bar{\nu}_i(2f'_{ik}P_R)(\slashed{k}+\slashed{p}+m_k)(h_{kj}P_R) \nu_j
%=-2\mu_Dv_D m_kf'_{ik}h_{kj}\overline{\nu_{Li}}\nu_{Rj}
%\end{eqnarray}
%\begin{eqnarray}
%\mathcal{M}&=&-i\frac{2\mu_Dv_D}{m_{s1}^2-m_{s2}^2}f'_{ik}h_{kj} m_k\bar{\nu}_{Li}\nu_{Rj}\int_0^1dx
%(I_0^{(2)}(m_{s1}^2)-I_0^{(2)}(m_{s2}^2))\nonumber\\
%&=&-\frac{1}{16\pi^2}\frac{2\mu_Dv_D}{m_{s1}^2-m_{s2}^2}f'_{ik}h_{kj} m_k\bar{\nu}_{Li}\nu_{Rj} \int_0^1 dx[\log(xm_k^2+(1-x)m_{s1}^2)-\log(xm_k^2+(1-x)m_{s2}^2)]\nonumber\\
%&=&-\frac{\mu_Dv_D}{8\pi^2m_{k}}f'_{ik}h_{kj}I(m_{s1}^2,m_{s2}^2,m_k^2)\bar{\nu}_{Li}\nu_{Rj}
%\end{eqnarray}
where $I(m_{s1}^2,m_{s2}^2,m_{\beta}^2)$ is defined as
\begin{equation}
I(m_{s1}^2,m_{s2}^2,m_{\beta}^2)=\frac{m_{\beta}^2}{m_{s1}^2-m_{s2}^2}[\frac{m_{s1}^2}{m_{s1}^2-m_{\beta}^2}\log\frac{m_{s1}^2}{m_{\beta}^2}
-\frac{m_{s2}^2}{m_{s2}^2-m_{\beta}^2}\log\frac{m_{s2}^2}{m_{\beta}^2}]\;.
\end{equation}
In the limit of $m_{s1,2}\gg m_k$, the neutrino mass matrix is proportional to
charged lepton mass, as given by
\begin{equation}
\label{numass}
\begin{split}
(M_\nu)_{\alpha i} &= -\frac{1}{8\pi^2}f_{\alpha\beta}h_{\beta i}m_{\beta}\frac{\mu f_a}{m_{s1}^2-m_{s2}^2}\log\frac{m_{s1}^2}{m_{s2}^2} \\
&\approx -\frac{1}{8\pi^2}f_{\alpha\beta}h_{\beta i}{\mu m_{\beta}\over m_a}\frac{\Lambda_\mathrm{QCD}^2}{m_{s1}^2-m_{s2}^2}\log\frac{m_{s1}^2}{m_{s2}^2}
= - Cf_{\alpha\beta}m_{\beta}h_{\beta i}.
\end{split}
\end{equation}
We have replaced the PQ-symmetry breaking scale by the axion mass and the QCD scale in Eq.~(\ref{numass})
\footnote{We should mention that quantum gravity effects do not respect global symmetries~\cite{Krauss:1988zc}, 
hence, the effective operator of Dirac mass receives an additional contribution suppressed by $\kappa\frac{L\nu^cHa}{M_{pl}}$ 
with $\kappa$ is the coefficient. If one requires this extra contribution is sub-dominant, say, less than $10^{-3}$ eV, $\kappa$ should 
be smaller than $10^{-7}$.}.
Note that there are similar construction discussed in Refs.~\cite{Nasri:2001ax,Kanemura:2011jj}.
However, our setup shows the interesting relation between the neutrino Dirac mass and QCD dynamics.
Furthermore, the axions can be the dark matter candidate constituting of $25\%$ of energy density in our universe. From the neutrino mass formula and scalar potential we can see that if one does not tune the couplings $h_{a1}$ and $h_{a2}$, then the masses of $s_1$ and $s_2$ should be of the order of $f_a$. Combining with the value of $\mu$ to be around the electroweak scale would directly lead to a small quantity of $C\approx 10^{-8}$, which means one can have the observed light neutrino masses with the Yukawa couplings $f,h$ at the order of $10^{-1}$. Although such a scenario can naturally provide a tiny neutrino mass without fine tuning the couplings, in what follows we still focus on the case in which both $m_{s1}$ and $m_{s2}$ are of the electroweak scale to provide richer phenomenological implications. In this case we can
see that the large value of $f_{a}$ lifts up the mass scale of Dirac neutrino masses and in order to keep the smallness of
$M_{\nu}$ a combination of suppressed factors will be needed such as the loop factor, the Yukawa couplings $f$, $h$,
the charged lepton chirality suppression, and the parameter $\mu$. Now we roughly estimate the scale of $\mu$ which is
the key parameter controlling the overall mass scale of Dirac neutrino masses and a investigation of parameter space will
be discussed in section~III. For $M_{\nu}\approx0.1$ eV, $m_{s_{1,2}} \sim \mathcal{O}(100 - 1000)$ GeV,
and $f_{a} \approx 10^{12}$ GeV, we have $f\sim10^{-3}$, $h\sim10^{-2}$, and $\mu\sim\mathcal{O}(1)$ keV. Let's
make two comments on the low scale $\mu$ and Dirac neutrino masses : {\bf 1.} We can explain the small $\mu$ by
implementing the Froggatt-Nielsen mechanism~\cite{CERN-TH-2519} with PQ-symmetry. For example, if we assign the
PQ quantum number of $a$ as $\frac{2}{n}$ in Table~\ref{model}, all the terms in the Lagrangian will not change except
the $\mu$-coupling in the potential. One can write the effective operator as
\begin{equation}
\frac{1}{\Lambda^{n-2}}s_1^{+}s_2^{-}a^{n} \xrightarrow[U(1)_{PQ} ~ breaking]{} (\frac{\langle a\rangle}{f_{a}})^{n-2}\langle a\rangle s_1^{+}s_2^{-}a,
\end{equation}
thus $\mu = (\frac{\langle a\rangle}{f_{a}})^{n-2}\langle a\rangle$ at low energy scale and can be tuned to a small quantity.
Here the PQ symmetry can be broken dynamically by some condensates of a new technicolor-like interaction
at high scale~\cite{Kim:1984pt,Choi:1985cb}. This mechanism can also apply to other dimensionless couplings of non-Hermitian terms such as the $h_5$ in the potential. {\bf 2.} The mass scale of Dirac neutrino masses is not necessarily small if
neutrinos are Majorana fermions. Although we consider neutrinos as the Dirac fermions, the main concern in this paper is
the origin of the Dirac neutrino mass, which in general does not forbidden the possibility that neutrinos are the Majorana
fermions. Therefore one can still have heavy right-handed Majorana masses as inspired by the grand unification theories and
obtain small neutrino masses through canonical seesaw mechanism. The goal in this paper is to point out that the Dirac neutrino
mass is generated by the PQ-symmetry breaking.

\section{Confronting with neutrino oscillation data and leptonic rare decays}
\label{sec:3}
From the standard formalism $M_{\nu}$ can be diagonalized by
%{\bf Case (b): }For $m_{s1}\simeq m_{s2}\simeq m_s$
%\begin{eqnarray}
%I(m_{s1}^2,m_{s2}^2,m_k^2)=I(m_s^2,m_k^2)
%=\frac{m_k^2}{m_s^2-m_k^2}+\frac{m_k^4}{(m_s^2-m_k^2)^2}\log\frac{m_k^2}{m_s^2}\,,
%\end{eqnarray}
%which is consistent with {\color{red} [Suematsu, Toma, Yoshida(2009)]}.
%{\bf Case (c): }Following the formula in which $m_s \simeq m_k$, e.g. fourth generation charged lepton, then $I(m_s^2,m_k^2)\simeq 1/2$, and
%\begin{eqnarray}
%(M_{\nu})_{ij}=-\frac{\mu_Dv_D}{16\pi^2}f'_{ik}h_{kj}m_k^{-1}\,.
%\end{eqnarray}
%Case (a) is discussed as follows.
%\begin{eqnarray}
%(M_\nu)_{ij}&=&-\frac{1}{8\pi^2}f'_{ik}h_{kj}{\mu_Dm_k\over m_a}\frac{\Lambda_\mathrm{QCD}^2}{m_{s1}^2-m_{s2}^2}\log\frac{m_{s1}^2}{m_{s2}^2}
%=Cf'_{ik}m_kh_{kj}
%\end{eqnarray}
\begin{equation}\label{massdiagonal}
M_{\nu_{diag}}=V_\mathrm{PMNS}^\dagger M_{\nu}V_R^{\nu\dagger}= - V_\mathrm{PMNS}^\dagger (CfM_{l_{diag}}h)V_R^{\nu\dagger},
\end{equation}
where $V_{\rm PMNS}$ is a unitary $3\times3$ Pontecorvo-Maki-Nakagava-Sakata (PMNS) matrix and $V_{R}$ is the
transformation matrix for right-handed neutrinos. For convenience we define $ FH=V_\mathrm{PMNS}M_{\nu_{diag}}$
with $H=M_{l_{diag}}(-hV_R^{\nu\dagger})$ and $F=Cf$. Due to the anti-symmetric nature of $f$ matrix, the lightest neutrino
is exactly massless in this model. Therefore, we can multiply a transformation matrix $A$ to both sides of the mass matrix
to reduce one row in the left hand side of Eq.~(\ref{massdiagonal}). One obtain
%Yukawa coupling in mass eigenstates
%\begin{eqnarray}
%\mathcal{L}_{Y+\mathrm{mass}}=[\overline{l_L}\hat M_ll_R(1+\frac{h_2^0}{v_2})+\overline{\nu_{mL}}\hat{M}_\nu \nu_{mR}
%+\overline{\nu_{mL}}(V_\mathrm{PMNS}^\dagger\hat M_l)l_Rh_2^++2\overline{\nu_L}f'(l_L)^cs_1^-+\overline{l_{R}^c}H\nu_{mR}s_2^+
%+\mathrm{h.c.}]
%\end{eqnarray}
%{\bf Method(a)} The relation can be expressed in terms of matrix elements
%\begin{eqnarray}
%\sum_{l'=e,\mu,\tau} F_{ll'}H_{l'j}=V_{ij}m_{\nu j}
%\end{eqnarray}
%\begin{eqnarray}
%{F_{e\mu }\over F_{\mu \tau}}={H_{\tau 1}\over H_{e1}}\,,
%{F_{e\tau }\over F_{\mu\tau }}=-{H_{\mu 1}\over H_{e1}}\,.\nonumber\\
%H_{\mu2}=-{H_{\mu1} H_{e2}\over H_{e1}}+{V_{\tau2}m_2\over F_{\tau\mu}}
%\end{eqnarray}
%{\bf Method (b)}
%From the relation
%\begin{eqnarray}
%FH=V_\mathrm{PMNS}\hat{M}_\nu
%\end{eqnarray}
\begin{equation}
F'H=AV_\mathrm{PMNS}M_{\nu_{diag}}\label{Eq_MassReduced},
\end{equation}
where
\begin{equation}
F'=AF=\left(\begin{array}{ccc}
-F_{e\mu}&0&F_{\mu\tau}\\
F_{e\tau}&F_{\mu\tau}&0\\
0&0&0\\
\end{array}\right),
A=\left(\begin{array}{ccc}
0&1&0\\
0&0&-1\\
F_{\mu\tau}&-F_{e\tau}&F_{e\mu}\\
\end{array}\right).
\end{equation}
Then we have
\begin{equation}
\begin{split}
&\left(\begin{array}{ccc}
-F_{e\mu}H_{e1}+F_{\mu\tau}H_{\tau1}&-F_{e\mu}H_{e2}+F_{\mu\tau}H_{\tau2}&-F_{e\mu}H_{e3}+F_{\mu\tau}H_{\tau3}\\
F_{e\tau}H_{e1}+F_{\mu\tau}H_{\mu1}&F_{e\tau}H_{e2}+F_{\mu\tau}H_{\mu2}&F_{e\tau}H_{e3}+F_{\mu\tau}H_{\mu3}\\
0&0&0\\
\end{array}\right) \\
&=
\left(\begin{array}{ccc}
m_{\nu1}V_{\mu1}&m_{\nu2}V_{\mu2}&m_{\nu3}V_{\mu3}\\
-m_{\nu1}V_{\tau1}&-m_{\nu2}V_{\tau2}&-m_{\nu3}V_{\tau3}\\
m_{\nu1}\sum_lV_{l1}F_l
&m_{\nu2}\sum_lV_{l2}F_l
&m_{\nu3}\sum_lV_{l3}F_l\\
\end{array}\right)
\end{split}
\end{equation}
with $F_l=(F_{\mu\tau},-F_{e\tau},F_{e\mu})$.

For the normal hierarchical spectrum with $m_{\nu_1} = 0$ we have the following relations :
%To satisfy the vanishing of the third row of Eq.~(\ref{Eq_MassReduced}), $A_{3l}\propto V_{l1}^*$, which means
\begin{equation}
F_{e\mu}={V_{\tau1}^*\over V_{e1}^*}F_{\mu\tau}\,,\,F_{e\tau}=-{V_{\mu1}^*\over V_{e1}^*}F_{\mu\tau}\,,
\end{equation}
and
%$m_{\nu_1}=0$, the vanishing of the first column leads to
\begin{equation}
H_{\mu1}={V_{\mu1}^*\over V_{e1}^*}H_{e1}\,,\,H_{\tau1}={V_{\tau1}^*\over V_{e1}^*}H_{e1}.
\end{equation}
The other terms give $H_{\mu2}$, $H_{\mu3}$, $H_{\tau2}$, $H_{\tau3}$ in terms of $F_{\mu \tau}$, $H_{e2}$ and $H_{e3}$
\begin{equation}
\begin{split}
H_{\mu2} &= {V_{\mu1}^*\over V_{e1}^*}H_{e2}-{m_{\nu_2}\over F_{\mu\tau}}V_{\tau2}\;,\;
H_{\tau2}={V_{\tau1}^*\over V_{e1}^*}H_{e2}+{m_{\nu_2}\over F_{\mu\tau}}V_{\mu2}\,,\, \\
H_{\mu3} &= {V_{\mu1}^*\over V_{e1}^*}H_{e3}-{m_{\nu_3}\over F_{\mu\tau}}V_{\tau3}\;,\;
H_{\tau3}={V_{\tau1}^*\over V_{e1}^*}H_{e3}+{m_{\nu_3}\over F_{\mu\tau}}V_{\mu3}\,.
\end{split}
\end{equation}
%The real $F$ matrix leads to the dependence of two phases in $V_\mathrm{PMNS}$ and only one irreducible phase remains.
Note that the requirement of real $F$ matrix make $V_\mathrm{PMNS}$ include two addition phases besides the ordinary irreducible one. Therefore, the model will not give a conclusive prediction for the Dirac CP phase in the neutrino sector at current stage.

Similarly for the inverted hierarchical spectrum, $m_{\nu_{3}} = 0$, we obtain
\begin{equation}
\begin{split}
F_{e\mu} &= {V_{\tau3}^*\over V_{e3}^*}F_{\mu\tau}\,,\,F_{e\tau}=-{V_{\mu3}^*\over V_{e3}^*}F_{\mu\tau}\,, \\
H_{\mu1} &= {V_{\mu3}^*\over V_{e3}^*}H_{e1}-{m_{\nu_1}\over F_{\mu\tau}}V_{\tau1}\,,\,
H_{\tau1} = {V_{\tau3}^*\over V_{e3}^*}H_{e1}+{m_{\nu_1}\over F_{\mu\tau}}V_{\mu1}, \\
H_{\mu2} &= {V_{\mu3}^*\over V_{e3}^*}H_{e2}-{m_{\nu_2}\over F_{\mu\tau}}V_{\tau2}\;,\;
H_{\tau2} = {V_{\mu3}^*\over V_{e3}^*}H_{e2}+{m_{\nu_2}\over F_{\mu\tau}}V_{\mu2}\,,\, \\
H_{\mu3} &= {V_{\mu3}^*\over V_{e3}^*}H_{e3}\;,\;
H_{\tau3} = {V_{\tau3}^*\over V_{e3}^*}H_{e3}\,.
\end{split}
\end{equation}
\begin{table}[tbp]
\centering
\begin{tabular}{|ccccc|}
  \hline
   $\sin^2\theta_{12}$ & $\sin^2 \theta_{23}$ &$\sin^2\theta_{13}$ &$\Delta m_{sun}^2$ &$\Delta m_{atm}^2$\\
   \hline
   $0.30\pm0.013$ & $0.41^{+0.037}_{-0.025}$ &$0.023\pm0.0023$&$(7.50\pm0.185)\times10^{-5}\mathrm{eV}^2$&$(2.47^{+0.069}_{-0.067})\times10^{-3}\mathrm{eV}^2$\\
  \hline
\end{tabular}
\caption{\label{input}Neutrino oscillation data.}
\end{table}
We will use the central values of the most recently global fitting of the neutrino oscillation
measurements~\cite{GonzalezGarcia:2012sz} (see Table~\ref{input}) in our analyses. In the meantime
the appearance of the new scalars will provide the extra contributions to the lepton flavor violation
processes. We investigate the parameter space of the model in terms of the constraints from leptonic
rare decays in the following. Before that it is worth mentioning that the size of parameter $C$ is proportional
to the factor $\mu f_a/(m_{s1}^2-m_{s2}^2)\log(m_{s1}^2/m_{s2}^2)$. If there exists a large hierarchy between
$m_{s1}$ and $m_{s2}$, say, $m_{s2}=k m_{s1}$ with $k$ being a large factor, $C$ is approximately inverse
proportional to $m_{s2}^2$. However, the positive mass eigenvalue condition $m_{s1}m_{s2}>\mu f_a$ will also
lead to the result that the upper bound of $C$ is proportional to $1/k$. In general $C$ can not be too small in
order to keep the perturbativity of the Yukawa coupling $f$ and $h$. So the largest allowed hierarchy between
$m_{s1}^2$ and $m_{s2}^2$ is around $k\simeq O(10^3)$. We will also discuss the implication to the parameter
space with a hierarchy scenario.
\subsection{$\mu\to e\gamma$}
\begin{figure}[tbp]
  \centering
  \includegraphics[width=75mm,height=60mm]{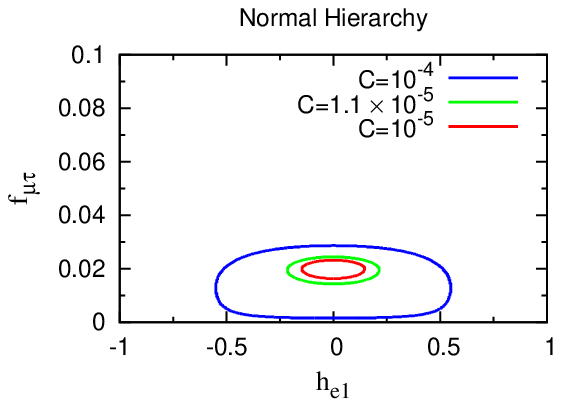}
  \includegraphics[width=75mm,height=60mm]{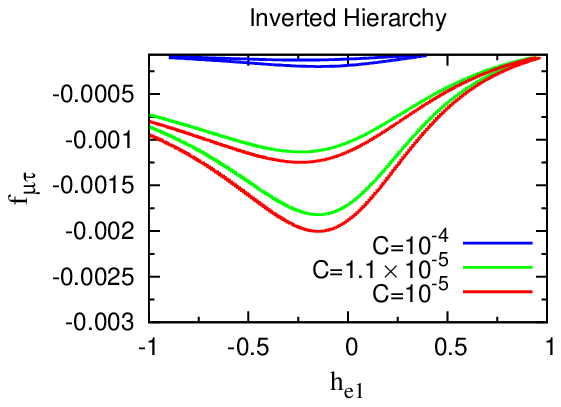}
  \caption{The allowed region in $h_{e1}-f_{\mu\tau}$ plane in (a) the normal hierarchy, and (b) the inverted hierarchy.
  Different curves are generated with different $C$ (blue: $C=10^{-4}$, green: $C=1.1\times10^{-5}$,
  red: $C=10^{-5}$). The other parameters are taken
  as $m_{s1}=m_{s2}= 500\mathrm{GeV}$, $h_{e2}=h_{e3}=0.5$.}\label{Fig_Couplings}
\end{figure}

The two relevant Yukawa interactions providing the flavor violations in charged lepton sector are given by
\begin{equation}
\mathcal{L} = -2f_{\alpha\beta}\overline{l_{L_{\alpha}}^c}\nu_{L_{\beta}}s_1 - h_{\alpha i}\overline{l_{R_{\alpha}}^c}\nu_{R_{i}}s_2+\mathrm{h.c.}.
\end{equation}
Without loss of generality here $\nu_{R_{i}}$ is the mass eigenstate and we absorb the mixing matrix $-V_{R}$ into the coupling $h$.
In $\mathrm{SM}+\nu_R$ the one-loop contribution is constrained stringently by the Glashow-Iliopoulos-Maiani (GIM)
mechanism. In this model the main contribution is the one-loop diagrams with photon emission attached to the charged scalars
$s_1^+$ or $s_2^+$ in the loop. Currently the latest result from MEG collaboration gives $B(\mu\rightarrow e\gamma)<2.4\times10^{-12}$~\cite{Adam:2011ch}. For $l_{\alpha}\to l_{\beta} \gamma$ the effective Lagrangian can be generally written in the form
\begin{equation}
\label{Eq_EMdipole}
\mathcal{L}=-{1\over2}\bar{l}_{\beta} \sigma_{\mu\nu}(\tilde A_R P_R+\tilde A_L P_L) l_{\alpha}F^{\mu\nu},
\end{equation}
where $\tilde{A}_{L,R}$ for this model is given by
\begin{equation}
\tilde A_R = \frac{(-1) e}{192\pi^2} (4\sum_\gamma f_{\beta\gamma}^*f_{\alpha\gamma}){m_{\alpha}\over m_{s1}^2} \quad {\rm and} \quad
\tilde A_L=\frac{(-1) e}{192\pi^2} (\sum_i h_{\beta i}^*h_{\alpha i}){m_{\alpha}\over m_{s2}^2}\,.
\end{equation}
Note that the limit $m_{\alpha}\ll m_{s1,2}$ have been applied to the above formula. The decay rate of
$l_{\alpha}\to l_{\beta}\gamma$ is $\Gamma(l_{\alpha}\to l_{\beta}\gamma)=(m_{\alpha}^3/16\pi)(1-m_{\beta}^2/m_{\alpha}^2)^3(|\tilde A_R|^2+|\tilde A_L|^2)$. One can compare it with the lepton three body decay width $\Gamma(l_{\alpha}\rightarrow l_{\beta}\nu_{\alpha}\bar\nu_{\beta})=G_F^2 m_{\alpha}^5/192\pi^3$. The branching ratio $\mu\rightarrow e\gamma$ in our model is obtained by
\begin{equation}
\text{B}(\mu\to e\gamma) \equiv \frac{\Gamma(\mu\to e\gamma)}{\Gamma(\mu\rightarrow e \nu\bar\nu)}=\frac{\alpha_e}{768\pi G_F^2}
\bigg({16\sum_\gamma|f_{e\gamma}|^2|f_{\mu\gamma}|^2\over m_{s1}^4}+{\sum_i|h_{ei}|^2|h_{\mu i}|^2\over m_{s2}^4}\bigg)\;.
\end{equation}
The results are given in Fig.~\ref{Fig_Couplings}. Here we consider that $f_{\mu\tau}$ is positive for the normal hierarchy
spectrum and negative for the inverted hierarchy spectrum, respectively. The allowed regions
are rather small and sensitive to the value of $C$. For example, the parameter space will shrink to zero if we take
$C = 0.8\times10^{-5}$ with the same inputs.

\subsection{$\mu-e$ conversion}
The one loop diagrams to $\mu-e$ conversion include photon penguin diagrams, Z penguin diagrams, and box diagrams.
Again the contribution from the diagrams involving W boson exchange is suppressed by GIM mechanism and the leading
contributions come from the penguin diagrams with the charged scalars $s_1^+$ and $s_2^+$ in the loop. In contrast, the
leading Z penguin contribution is suppressed by the light charged scalars. Therefore only photon penguin needs to be taken into
account. The corresponding effective Lagrangian is given by
\begin{equation}
\label{Eq_mueconversion}
\mathcal{L} = -\frac{G_F}{\sqrt{2}}\frac{s_W^2}{36\pi^2}m_W^2 \bar{e}\gamma_\mu (\sum_\alpha4f_{e\alpha}^*f_{\mu\alpha}\frac{1}{m_{s1}^2}P_L + \sum_ih_{ei}^*h_{\mu i}\frac{1}{m_{s2}^2}P_R)\mu\sum_q Q_q\bar{q}\gamma^\mu q\,.
\end{equation}
In the above we used the shorthand notation $s_{W} \equiv \sin{\theta_{W}}$.
The $\mu-e$ conversion with nucleon in atom has been calculated in details in Ref.~\cite{Kitano:2002mt},
and we will adopt their notation in what follows. The general interactions associated with $\mu-e$ conversion
are written as%~\cite{Kitano:2002mt}
\begin{equation}
\begin{split}
\mathcal{L}_\mathrm{eff} &= -{4G_Fe\over\sqrt{2}}[m_\mu\bar{e}\sigma^{\mu\nu}(A_RP_R+A_LP_L)\mu F_{\mu\nu}+\mathrm{h.c.}] \\
& -{G_F\over\sqrt{2}}[\bar e(g_{LS(q)}P_R+g_{RS(q)}P_L)\mu \bar q q
+\bar e(g_{LP(q)}P_R+g_{RP(q)}P_L)\mu\bar q\gamma_5q+\mathrm{h.c.}] \\
& -{G_F\over\sqrt{2}}[\bar e(g_{LV(q)}\gamma^\mu P_L+g_{RV(q)}\gamma^\mu P_R)\mu \bar q\gamma_\mu q
+\bar e(g_{LA(q)}\gamma^\mu P_L+g_{RA(q)}\gamma^\mu P_R)\mu \bar q\gamma_\mu\gamma_5 q+\mathrm{h.c.}] \\
& -{G_F\over\sqrt{2}}\bigg[{1\over2}\bar e(g_{LT(q)}\sigma^{\mu\nu}P_R+g_{RT(q)}\sigma^{\mu\nu}P_L)\mu\bar q\sigma_{\mu\nu}q+\mathrm{h.c.}\bigg]\,,
\end{split}
\end{equation}
where $A_{R,L}$ is related to the dipole interaction with photon, and $g_{S,P,V,A,T}$ indicate scalar, pseudoscalar, vector, axial
vector and tensor couplings, respectively. Comparing the above formula with Eq.~(\ref{Eq_mueconversion}), we have
\begin{subequations}
\begin{align}
g_{LV}^{(q)} &= \frac{Q_qs_W^2}{36\pi^2}m_W^2(\sum_\alpha 4f_{e\alpha}^*f_{\mu\alpha}\frac{1}{m_{s1}^2})\,,\,
g_{RV}^{(q)}=\frac{Q_qs_W^2}{36\pi^2}m_W^2(\sum_ih_{ei}^*h_{\mu i}\frac{1}{m_{s2}^2})\;,\\
A_R &= \frac{(-1) }{192\pi^2g_2^2} (\sum_\alpha 4f_{e\alpha}^*f_{\mu\alpha}{m_W^2\over m_{s1}^2})\;,
A_L=\frac{(-1) }{192\pi^2g_2^2} (\sum_ih_{ei}^*h_{\mu i}{m_W^2\over m_{s2}^2}),
\end{align}
\end{subequations}
and other couplings vanish. The rate of $\mu-e$ conversion with nucleons in atom $A$ is usually normalized to the rate of
muon capture by $A$. The conversion-to-capture ratio can be derived in the form
\begin{equation}
B_{\mu\rightarrow e}^A={2G_F^2m_\mu^5\over\Gamma_\mathrm{capt}^A}(|eA_RD+\tilde{g}_{LV}^{(p)}V^{(p)}+\tilde{g}_{LV}^{(n)}V^{(n)}|^2
+|eA_LD+\tilde{g}_{RV}^{(p)}V^{(p)}+\tilde{g}_{RV}^{(n)}V^{(n)}|^2)\;
\end{equation}
with $\tilde g_{L,RV}^{(p)}=2g_{L,RV}^{(u)}+g_{L,RV}^{(d)}$
and $\tilde g_{L,RV}^{(n)}=g_{L,RV}^{(u)}+2g_{L,RV}^{(d)}$. $D$ and $V^{(p,n)}$
are overlapped functions which can be found in Ref.~\cite{Kitano:2002mt}. The constraints from
experimental results for $\mu-e$ conversion in different
nuclei~\cite{Bertl:2006up,Badertscher:sulfer,Badertscher:1981ay,Dohmen:1993mp,Honecker:1996zf} are
weaker than what's given by $\mu\rightarrow e\gamma$.

\subsection{$\mu\rightarrow 3e$}
Penguin diagrams and Box diagrams contribute to this process. The experimental upper bounds
is $B(\mu\rightarrow e\bar{e}e)<10^{-12}$~\cite{Bellgardt:1987du}. SM with right-handed neutrino singlets contribution
to this processes is suppressed by the neutrino mass. The corresponding one loop diagrams in this model for $\mu\rightarrow3e$
are similar to those for $\mu-e$ conversion, with the quarks replaced by electrons. In general the effective Lagrangian
for $\mu\rightarrow e\bar ee$ is
\begin{equation}
\begin{split}
\mathcal{L}(\mu\rightarrow e\bar ee) &= -{4G_F\over\sqrt{2}}[\bar e\gamma_\mu e{q_\nu\over q^2}\bar ei\sigma^{\mu\nu}(8\pi\alpha_e m_\mu )(A_R P_R+A_L P_L)\mu \\
&+ \bar e\gamma^\mu(a_LP_L+a_RP_R)e\bar e\gamma_\mu P_L\mu+\bar e\gamma^\mu(b_LP_L+b_RP_R)e\bar e\gamma_\mu P_R\mu]\,.
\end{split}
\end{equation}
The branching ratio can be calculated as
\begin{equation}
\begin{split}
B(\mu\rightarrow e\bar ee) &\simeq (|a_{R}|^2+|b_{L}|^2)+2(|a_{L}|^2+|b_{R}|^2)
-32\pi \alpha_e\mathrm{Re}(A_R(a_{R}+2a_{L})+A_L(b_{L}+2b_{R})) \\
& + 256\pi^2\alpha_e^2(|A_R|^2+|A_L|^2)\bigg(4\ln{m_\mu\over m_e}-{11\over2}\bigg)\,,
\end{split}
\end{equation}
where the parameters $a_{L,R}$ and $b_{L,R}$ are given by
\begin{equation}
a_{L,R} = -\frac{s_W^2}{144\pi^2}m_W^2(\sum_\alpha 4f_{e\alpha}^*f_{\mu\alpha}\frac{1}{m_{s1}^2}) \quad  {\rm and} \quad
b_{L,R} = -\frac{s_W^2}{144\pi^2}m_W^2(\sum_ih_{ei}^*h_{\mu i}\frac{1}{m_{s2}^2})\,
\end{equation}
from the photon penguin diagrams, while for the box diagrams the leading order of $a^{\rm box}_{R}$ and $b^{\rm box}_{L}$ vanishing,
we have
\begin{subequations}
\begin{align}
a^{\rm box}_{L} = {m_W^2\over32\pi^2g_2^2m_{s1}^2}|\sum_{\alpha} 4f_{\mu\alpha}f_{e\alpha}^*||\sum_{\alpha'} 4f_{e\alpha'}^*f_{e\alpha'}|
\end{align}
and
\begin{equation}
b^{\rm box}_{R} = {m_W^2\over32\pi^2g_2^2m_{s2}^2}|\sum_i h_{\mu i}h_{ei}^*||\sum_{i'} h_{ei'}^*h_{ei'}|
\end{equation}
\end{subequations}
respectively. As shown in Fig.~\ref{Fig_Couplings} the Yukawa coupling $f$ is in the range of $\cal{O}$($10^{-2}$) to $\cal{O}$($10^{-3}$);
we can safely ignore the box diagram contributions in $\mu\rightarrow3e$ decay. Therefore, the parameter space is
looser to those of $\mu\rightarrow e\gamma$.

\begin{figure}[tbp]
  \centering
  \includegraphics[width=75mm,height=60mm]{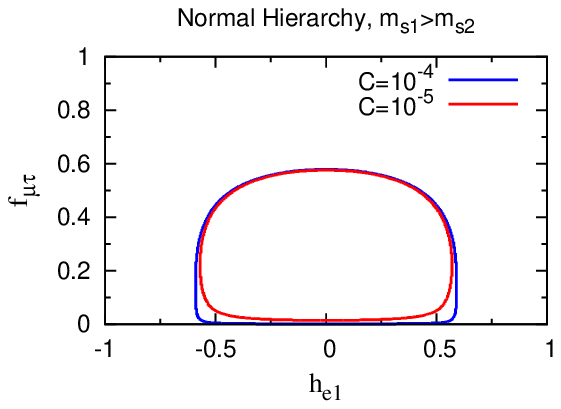}
  \includegraphics[width=75mm,height=60mm]{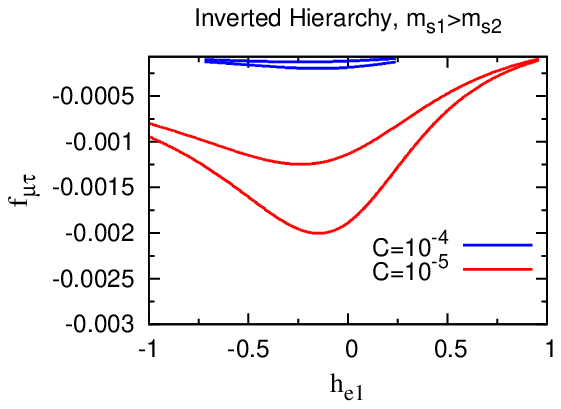}
  \caption{The allowed region in $h_{e1}-f_{\mu\tau}$ plane for the case of $m_{s1} = 20m_{s2}$
  in normal hierarchy and inverted hierarchy neutrino mass spectrum respectively. Blue line corresponds to $C=10^{-4}$
  and Red line represents $C=10^{-5}$.}\label{Fig_HIcouplings1}
\end{figure}

\begin{figure}[tbp]
  \centering
  \includegraphics[width=75mm,height=60mm]{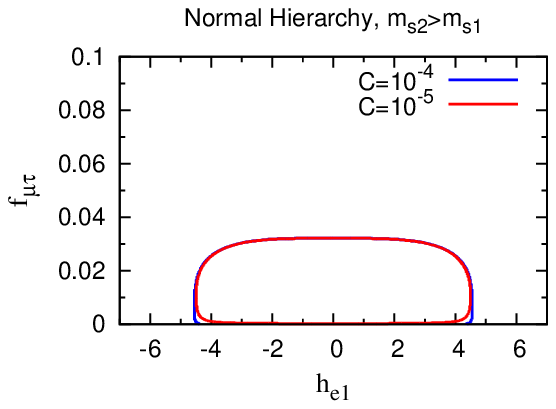}
  \includegraphics[width=75mm,height=60mm]{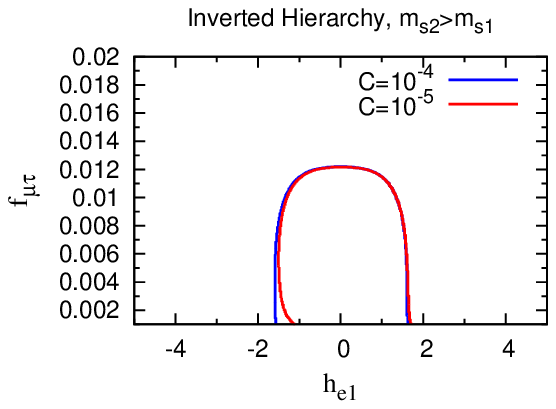}
  \caption{The allowed region in $h_{e1}-f_{\mu\tau}$ plane for the case of $m_{s2} = 5m_{s1}$
  in normal hierarchy and inverted hierarchy neutrino mass spectrum respectively.
  Blue line corresponds to $C=10^{-4}$ and Red line represents $C=10^{-5}$.}\label{Fig_HIcouplings2}
\end{figure}

In the limit of mass hierarchy between $m_{s1}$ and $m_{s2}$, that is, $m_{s1} > m_{s2}$ or $m_{s2} > m_{s1}$ cases,
the parameter space for Yukawa couplings $h$ and $f$ in both normal hierarchy and inverted hierarchy neutrino mass
spectrum respectively are shown in Fig.~\ref{Fig_HIcouplings1} and Fig.~\ref{Fig_HIcouplings2}.
Here we take $m_{s1} = 20m_{s2}$ and $m_{s2} = 5m_{s1}$ as the reference points. In these cases, we are only able
to give a severe constraint on one of the couplings, $h$ or $f$, and we illustrate our results by taking $500$ GeV mass
scale to the lighter scalar field.

\section{Phenomenology}
\label{sec:4}
\subsection{electromagnetic moments of leptons}
\subsubsection{muon $g-2$}
The anomalous magnetic moment of the muon is an observable as a precision test to the SM. The
current experimental results~\cite{Bennett:2006fi,pdg2012} reported that the muon $g-2$ deviation
from the SM prediction is $a_{\mu}(\rm{exp}) - a_{\mu}(\rm{SM}) = (287\pm63\pm49)\times10^{-11}$,
which indicates some new physics contribution.
%The anomalous magnetic dipole moment of charged lepton is given by {\color{red}Hisano {\it et al.}(1996)}
%\begin{eqnarray}
%\mathcal{M}=\frac{e}{2m_{\alpha}}a_{\alpha}\,\epsilon^{*\mu}\bar{l_{\alpha}} i\sigma_{\mu\nu}q^\nu l_{\alpha}\;.
%\end{eqnarray}
%One can expressed the anomalous magnetic dipole moment, $a_{\alpha}$, in terms of $\tilde A_{R,L}$ as
%\begin{eqnarray}
%a_{\alpha}=\frac{g_{\alpha}-2}{2}=\frac{2m_{\alpha}}{e}\frac{\tilde A_R+\tilde A_L}{2}\;,
%\end{eqnarray}
%where $\tilde A_{R,L}$ are the amplitudes shown in Eq.~(\ref{Eq_EMdipole}) with $\alpha=\beta$.
Since our scenario is essentially a two Higgs doublet model (2HDM) plus two singly charged scalars
$s^{\pm}_{1}$ and $s^{\pm}_{2}$. The anomalous $g-2$ of muon in this model is given by
\begin{equation}
\begin{split}
\Delta a_\mu= -{m_\mu^2\over 96\pi^2} & [4\sum_\alpha |f_{\mu\alpha}|^2\frac{1}{m_{s1}^2}
+ \sum_i |h_{\mu i}|^2\frac{1}{m_{s2}^2} \\
& + {m_\mu^2\over v^2}({14c_{\beta-\alpha}^2\over m_{h}^2\cos^2\beta }
+ {14s_{\beta-\alpha}^2\over m_{H}^2\cos^2\beta }-{22 \tan^2\beta\over m_{A}^2}+{\tan^2\beta\over m_{H^+}^2})]\;,
\end{split}
\end{equation}
where $h$ and $H$ are scalar particles with $h$ being the SM-like Higgs, as well as the pseudo-scalar field $A$ and
the charged scalars $H^{\pm}$. $\alpha$, $\beta$ are the mixing angles
($s_{\beta-\alpha} \equiv \sin{(\beta-\alpha)}$ and $c_{\beta-\alpha} \equiv \cos{(\beta-\alpha)}$) defined in the
usual 2HDM (see e.g.~\cite{Branco:2011iw, Chang:2012ta} and references within). %The terms proportional to $m_\mu^2/v^2$ can is negligible. %For $m_{s1}=m_{s2}=1\mathrm{TeV}$, $f_{\mu\tau}=0.0015$, $h_{e2}=h_{e3}=10^{-3}$, it leads to $\Delta a_\mu=-1.2\times10^{-16}$.
Comparing with the current experimental result, our model gives a rather small contribution to $\Delta{a_{\mu}}$ but
still within the experimental errors.

\subsubsection{magnetic moments of neutrinos}
Neutrinos can have magnetic moment when they are massive. The present upper bound of neutrino
magnetic moment from experiments is $\mu_\nu<3.2\times10^{-11}\mu_B$~\cite{Beda:2009kx}.
The ordinary leading order contribution in SM including right-handed Dirac neutrino is the exchange
of $W$, which is around $(3.2\times10^{-19}\mu_B)(m_{\nu_i}/\mathrm{eV})$~\cite{Fujikawa:1980yx}
with the Bohr magneton $\mu_B\equiv e/2m_e$. In this model, the main new contribution comes from the
mixing of $s^{+}_{1,2}$ in the loop since the same diagram also generates neutrino masses, given by
\begin{equation}
|\mu_\nu^{s}| = 2m_ie(\log{m_{s2}^2\over m_{s1}^2})^{-1}({1\over m_{s1}^2}-{1\over m_{s2}^2})\approx 10^{-19}\mu_B.
\end{equation}
Note that it only depends on the neutrino and scalar masses, and this contribution is comparable with that associated
with W exchange. It is understandable that without imposing a symmetry~\cite{Voloshin:1987qy} or employing a spin
suppression mechanism to $m_{\nu}$~\cite{Barr:1990um}, the generic size of the Dirac neutrino magnetic moment can
not excess $10^{-14}\mu_{B}$~\cite{Bell:2005kz}. The contributions from other charged scalars such as $H^+$, $s^{+}_{1}$,
and $s_{2}^+$ in the loop are less than $\mathcal{O}(10^{-25})\mu_B$ as a result that they do not contribute to neutrino masses.

\subsubsection{{\rm EDM} of leptons}
Since there is a physical CP phase among the Yukawa couplings $f$ and $h$, in general, we have the new contributions to
the electric dipole moment of charged leptons. The exchange of singly charged scalars gives the EDM to the charged
leptons at one-loop level as
%For the tree-level neutral and charged Higgs Yukawa interaction
%\begin{eqnarray}
%\mathcal{L}_Y=\bar{\ell}_i(N_{Rij}P_R+N_{Lij}P_L)\ell_j H^0
%+\bar{\ell}_i(C_{Rij}P_R+C_{Lij}P_L)\nu_j H^-+\mathrm{h.c.}\,
%\end{eqnarray}
%Ref.\cite{Ibrahim:1997gj} provide some formalism for EDM of loop calculation, which is given by
%\begin{eqnarray}
%d_l^n=-\frac{e}{(4\pi)^2}\text{Im}(N_{Lik}N_{Rik}^*){1\over m_{H^0}}A\left(\frac{m_k^2}{m_{H^0}^2}\right),\nonumber\\
%d_l^c=-\frac{e}{(4\pi)^2}\text{Im}(C_{Llk}C_{Rlk}^*){1\over m_{H^-}}B\left(\frac{m_{\nu_k}^2}{m_{H^-}^2}\right),
%\end{eqnarray}
%with
%\begin{eqnarray}
%A(x)=\frac{\sqrt{x}}{2(1-x)^2}\left(3-x+\frac{2\ln(x)}{1-x}\right);\nonumber\\
%B(x)=\frac{\sqrt{x}}{2(1-x)^2}\left(1+x+\frac{2x\ln(x)}{1-x}\right),
%\end{eqnarray}
%where $d_l^n$ and $d_l^c$ are the EDM contribution from exchange of neutral boson and charged boson in one loop, respectively. In this model, the lepton EDM is given by
%\begin{eqnarray}
%d_l&=&-\frac{e}{(4\pi)^2}\text{Im}(C_{1lk}^LC_{2lk}^{R*}+C_{2lk}^LC_{1lk}^{R*})
%\frac{\mu f_a}{m_{S_1^-}^2-m_{S_2^-}^2}
%[{1\over m_{S_1^-}}B\left(\frac{m_\nu^2}{m_{S_1^-}^2}\right)-{1\over m_{S_2^-}}B\left(\frac{m_\nu^2}{m_{S_2^-}^2}\right)]\;,\nonumber\\
%\end{eqnarray}
%where $C_{1R}=2f^*$, $C_{2L}=h^*$, then we have
\begin{equation}
d_l = -\frac{e}{(4\pi)^2}\sum_i\mathrm{Im}(2h_{li}^*(fV)_{li})
\frac{\mu f_a}{m_{s1}^2 - m_{s2}^2}
[{1\over m_{s1}}B\left(\frac{m_{\nu_i}^2}{m_{s1}^2}\right)-{1\over m_{s2}}B\left(\frac{m_{\nu_i}^2}{m_{s2}^2}\right)]\;,
\end{equation}
with the function $B(x)$ defined by
\begin{equation}
B(x)=\frac{\sqrt{x}}{2(1-x)^2}\left(1+x+\frac{2x\ln(x)}{1-x}\right).
\end{equation}
Therefore we conclude that the extra contribution is smaller than $\mathcal{O}({m_\nu^2\over m_l m_s^2})$,
which means $|d_e|\lesssim10^{-35}e\,\mathrm{cm}$. Similarly, the neutrino EDM generated by $s^+_{1,2}$
mixing in the loop gives $|d_\nu|\lesssim10^{-26}e\,\mathrm{cm}$. Though the new contributions to $d_e$ and $d_\nu$ are nonzero, both of them are unobservably small.

\subsection{dark matter}
It is known that the axion field can be a dark matter candidate constituting a significant fraction of energy
density in our universe. For the purpose of completion we briefly review some aspects of this scenario in this subsection.
The properties of the invisible axion are determined by the breaking scale $f_{a}$ where its mass and the
interactions are inverse proportional to $f_{a}$. Hence the invisible axion is a very light, very
weakly interacted and very long-lived particle. Axions with mass in the range of $10^{-5}-10^{-6}$ eV were produced
during the QCD phase transition with the average momentum of order the Hubble expansion rate
($\sim3\times10^{-9}$ eV) at this epoch and hence are cold dark matter (CDM). Their number density is provided by
\begin{equation}
\Omega_{a} \simeq \frac{1}{2}(\frac{0.6\times10^{-5}~\rm{eV}}{m_{a}})^{\frac{7}{6}}(\frac{0.7}{h})^2,
\end{equation}
where $h$ is the current Hubble expansion rate in units of 100$\rm{km s^{-1} Mpc^{-1}}$. Here we assume the ratio
of the axion number density to the entropy density is constant since produced and the contribution from
topological defects decay is negligible. Recently it was pointed out that the CDM axions would form a Bose-Einstein
condensate due to their gravitational interactions~\cite{Sikivie:2009qn}. Furthermore the rethermalization process is so
fast that the lowest energy state of the degenerate axion gas consists of a nonzero angular momentum. As a result
a "caustic ring" structure may form in the inner galactic halo~\cite{Erken:2011dz,Sikivie:1997ng,Sikivie:1999jv}.
The feature would make the axion a different dark matter from other CDM candidates, and we refer readers to the references~\cite{Sikivie:2009qn,Erken:2011dz,Sikivie:1997ng,Sikivie:1999jv} for details.
%For CDM with $m_D< 15\mathrm{GeV}$ and $m_H\simeq125\mathrm{GeV}$, the saturation of the relic abundance leads to $\mathrm{Br}(h\rightarrow DD)\simeq1${\color{red}(He and Jusak, 2011)}\\

%$f_{D}(H_1^\dagger H_1)|D|^2$, $g_{D}(H_2^\dagger H_2)|D|^2$, $h_{5}(H_1^\dagger H_2)D^2$

\subsection{$h\rightarrow \gamma\gamma$}
We closing our discussion on phenomenology with investigating the LHC Higgs results. Both
ATLAS~\cite{:2012gk} and CMS~\cite{:2012gu} have announced the discovery of a new boson at a
mass of 125 GeV which is consistent with the SM Higgs boson via the combined analyses of
the $\gamma\gamma$ and $ZZ$ channels. However, the precise values of both production cross sections
and decay branch ratios of the new resonance need to be measured to compare with those predictions from the
SM. It was pointed out that the branching ratio of Higgs decay into two photons has excess about $1.56\pm0.43$ and
$1.9\pm0.5$ times than the SM prediction in both CMS~\cite{:2012gu} and ATLAS~\cite{:2012gk} collaborations data in 2012. A updated results can be found in \cite{ATLAS_NOTE_2013_034} and \cite{moriond2}. 
Although the deviation is still within the SM expectations at $2\sigma$ level, one may consider whether there are new
physics effects (see e.g.~\cite{Cacciapaglia:2009ky,Cao:2011pg,Ellwanger:2011aa,Barger:2012hv,Azatov:2012bz,
Giardino:2012ww,Carena:2012xa,Christensen:2012ei,Carena:2012gp,Chang:2012ta,Chiang:2012qz} and references therein)
In particular, new scalars particles have been widely treated as possible sources. We study the implications
of Higgs to di-phonton and Higgs to Z-photon decay channels in our model. For simplicity we just consider one singly charged
scalar $s^{\pm}$ and omit the subscript in this subsection. %Since our scenario is essentially a two Higgs
%doublet model (2HDM) plus two singly charged scalars $s^{\pm}_{1}$ and $s^{\pm}_{2}$.
The SM Higgs production cross section is modified by the additional doublet scalar in our scenario~\cite{Chang:2012ta},
\begin{equation}
\sigma_0={G_F\alpha_s^2\over128\sqrt{2}\pi}\bigg|{1\over2}(s_\alpha+{c_\alpha\over \tan\beta})A_{1/2}(\tau_t)
+{1\over2}(s_\alpha-c_\alpha\tan\beta)A_{1/2}(\tau_b)\bigg|^2\;.
\end{equation}
Notice that the bottom quark contribution is not negligible due the enhancement of large $\tan{\beta}$ and
$A_{1/2}(\tau) = 2[\tau+(\tau-1)f(\tau)]\tau^{-2}$ with $\tau_{i} = \frac{M^2_{h}}{4M^2_{i}}$. The function $f(\tau)$ is defined by
\begin{equation}
f(\tau)=
\bigg\{\begin{array}{l}
(\sin^{-1}\sqrt{\tau})^2\hspace{60pt}, \tau\leq 1\\
-{1\over4}[\log{1+\sqrt{1-\tau^{-1}}\over1-\sqrt{1-\tau^{-1}}}-i\pi]^2\quad ,\,\tau>1\,.
\end{array}
\end{equation}
In type-II 2HDM the decay rate of $h\rightarrow \gamma\gamma$ is given by
\begin{equation}
\begin{split}
\Gamma_{\gamma\gamma} &= {G_F\alpha^2 m_h^3\over 128\sqrt{2}\pi^3}|(s_\alpha+{c_\alpha\over\tan\beta}){4\over3}A_{1/2}(\tau_t)
+(s_\alpha-{c_\alpha\tan\beta}){1\over3}A_{1/2}(\tau_b) \\
& + s_\alpha A_1(\tau_W) + \lambda'A_0(\tau_{H^{+}}) + \lambda A_{0}(\tau_{s^{+}})|^2,
\end{split}
\end{equation}
where $A_{1}(\tau) = -[2\tau^2+3\tau+3(2\tau-1)f(\tau)]\tau^{-2}$,
and $A_{0}(\tau) = -[\tau-f(\tau)]\tau^{-2}$. The coefficients $\lambda$ and $\lambda'$ are defined as
$\lambda={v\mu_{Hs^+s^-}/(2m_s^2)}$ and $\lambda'={v\mu_{HH^+H^-}/(2m_{H^+}^2)}$, where $\mu_{Hs^+s^-}$
and $\mu_{HH^+H^-}$ are the related trilinear couplings in the Higgs potential.
One can see that the effects of doublet scalar charged component $H^{+}$ and the singly charged singlet $s^{+}$ are
indistinguishable. The reason is the lack of knowledge of scalar potential. For $h \rightarrow Z\gamma$ channel the
decay width is written as~\cite{Chen:2013vi}
\begin{equation}\label{Zr}
\begin{split}
\Gamma_{Z\gamma} &= {G_F^2m_W^2\alpha m_h^3\over 64\pi^4c_W^2}(1-m_Z^2/m_H^2)^3\times \\
& \bigg|(s_\alpha+{c_\alpha\over\tan\beta}){({1\over2}-{4\over 3}s_W^2)}A_{1/2}(\eta_t,\kappa_t)
+ (s_\alpha-{c_\alpha\tan\beta}){({1\over4}-{1\over3}s_W^2)}A_{1/2}(\eta_b,\kappa_b) \\
& + s_\alpha c_W^2A_1(\eta_W,\kappa_W)+({1\over2}-s_W^2)\lambda'A_0(\eta_{H^+},\kappa_{H^+})+(-
s_W^2)\lambda A_0(\eta_{s^+},\kappa_{s^+})
\bigg|^2, \\
\end{split}
\end{equation}
where $A_{1/2}(\eta,\kappa) = -4(I_1(\eta,\kappa)-I_2(\eta,\kappa))$,
$A_1(\eta,\kappa) = -4(4-{4\over\kappa})I_2(\eta,\kappa)
- [(1+{2\over\eta})({4\over\kappa}-1)-(5+{2\over \eta})]I_1(\eta,\kappa)$,
and $A_0(\eta,\kappa) = 2I_1(\eta,\kappa)$ with the
parameters $\eta_{i} = \frac{4M^2_{i}}{M^2_{h}}$ and $\kappa_{i} = \frac{4M^2_{i}}{M^2_{Z}}$.
Functions $I_{1}(\eta,\kappa)$ and $I_{2}(\eta,\kappa)$ are given as
\begin{equation}
I_1(\eta,\kappa) = {\eta\kappa\over 2(\eta - \kappa)}
+ {\eta^2\kappa^2\over 2(\eta - \kappa)^2}[f(\eta^{-1})-f(\kappa^{-1})]
+ {\eta^2\kappa\over (\eta - \kappa)^2}[g(\eta^{-1})-g(\kappa^{-1})]
\end{equation}
and
\begin{equation}
I_2(\eta,\kappa) = -{\eta\kappa\over2(\eta - \kappa)}[f(\eta^{-1})-f(\kappa^{-1})]
\end{equation}
with
\begin{equation}
g(x)=
\bigg\{\begin{array}{ll}
\sqrt{x^{-1}-1}(\sin^{-1}\sqrt{x})\hspace{60pt}&, x\leq 1\\
{\sqrt{1-x^{-1}}\over 2}[\log{1+\sqrt{1-x^{-1}}\over1-\sqrt{1-x^{-1}}}-i\pi]\quad &,\,x>1
\end{array}
\end{equation}
\begin{figure}[tbp]
  \centering
  \includegraphics[width=75mm,height=60mm]{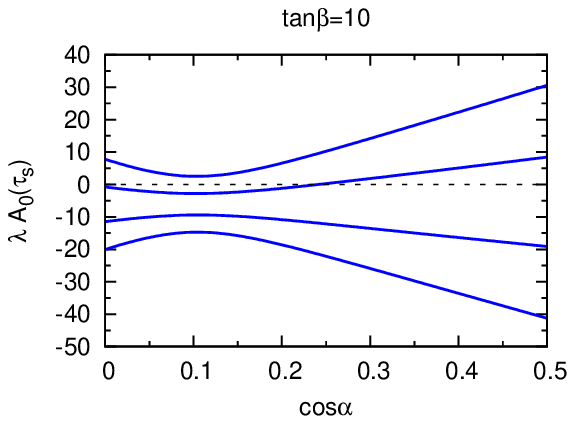}
  \includegraphics[width=75mm,height=60mm]{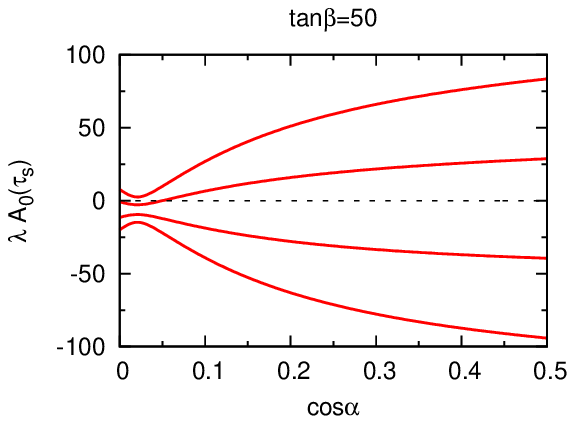}
  \caption{Solutions of $s^{+}$ effects in terms of $\cos{\alpha}$ by assuming $A_{0}(\eta,\kappa)/A_{0}(\tau) = 1.00$
  with $R_{\gamma\gamma} = \frac{\sigma_{\gamma\gamma}}{\sigma_{\gamma\gamma_{SM}}} = 1.5$
  and $R_{Z\gamma} = \frac{\sigma_{Z\gamma}}{\sigma_{Z\gamma_{SM}}} = 1$. $\sigma_{ii}$ and $\sigma_{ii_{SM}}$ are
the production cross sections for the Higgs to $ii$ channel in our model and SM respectively.}\label{Fig_s1}
\end{figure}
\begin{figure}[tbp]
  \centering
  \includegraphics[width=75mm,height=60mm]{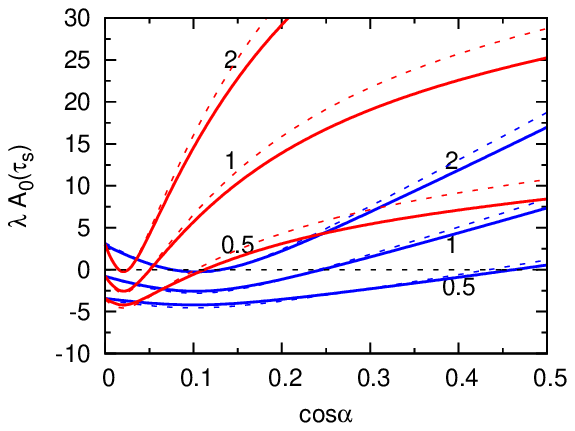}
  \includegraphics[width=75mm,height=60mm]{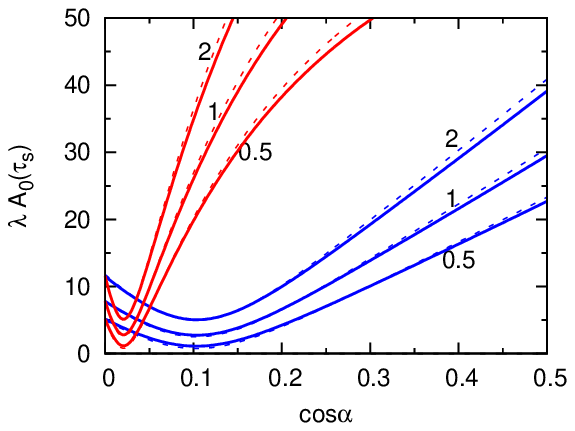}
  \includegraphics[width=75mm,height=60mm]{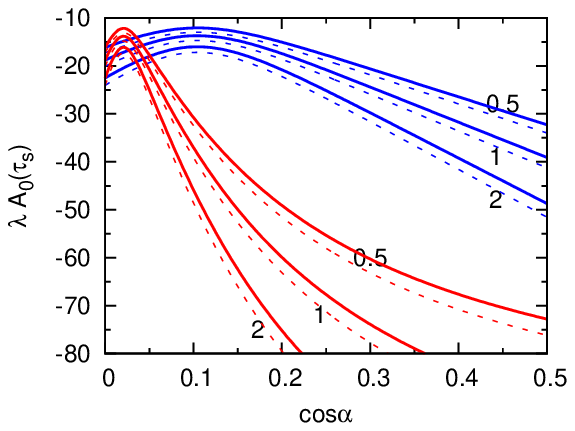}
  \includegraphics[width=75mm,height=60mm]{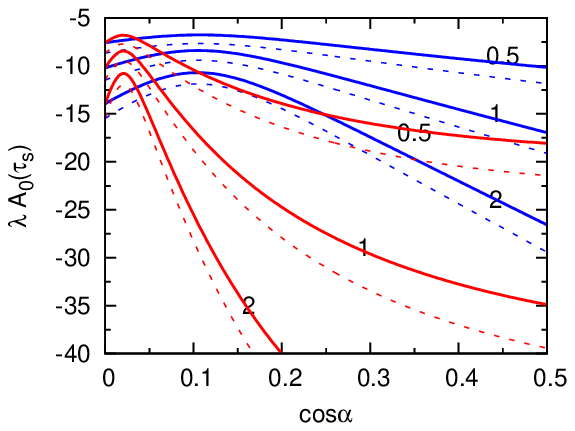}
  \caption{Effects of singly charged singlet scalar with values of $R_{Z\gamma} = 0.5, 1, 2$ as indicated on each curves, and $A_{0}(\eta,\kappa)/A_{0}(\tau) = 1.07$ (solid lines), $1.00$ (dashed lines) respectively for the fixed value $R_{\gamma\gamma} = 1.5$. Blue lines correspond to $\tan{\beta} = 10$ and Red lines represent $\tan{\beta} = 50$. Note that each figure corresponds to one of the lines exhibited in Fig.~\ref{Fig_s1}.}\label{Fig_s2}
\end{figure}
respectively. Again we see from the Eq.~(\ref{Zr}) that the singly charged singlet effect is hidden in the 2HDM.
In order to extract the information of $s^{+}$ from the decays we found the ratio of $A_{0}(\eta,\kappa)/A_{0}(\tau)$ lies in
the range of $1.00 - 1.07$ with charged scalar masses above $100$ GeV. We then take the ratio as a constant
and subtract the $H^{+}$ contributions in $\Gamma_{\gamma\gamma}$ and $\Gamma_{Z\gamma}$. Other SM Higgs decaying channels are calculated from HDECAY~\cite{Djouadi:1997yw}. The results
of the $s^{+}$ effects in terms of the parameter $\cos{\alpha}$ with $\tan{\beta} = 10$, $50$ are shown in Fig.~\ref{Fig_s1}.
Here we take the values
$R_{\gamma\gamma} = \frac{\sigma_{\gamma\gamma}}{\sigma_{\gamma\gamma_{SM}}} = 1.5$
and $R_{Z\gamma} = \frac{\sigma_{Z\gamma}}{\sigma_{Z\gamma_{SM}}} = 1$ with $\sigma_{ii}$ and $\sigma_{ii_{SM}}$ being
the production cross sections for the Higgs to $ii$ channel in our model and SM respectively. Since the new physics effects
can have both constructive and destructive interferences with the SM $W^{\pm}W^{\pm}$ amplitude, there are four lines exist
in Fig.~\ref{Fig_s1}. The results of $s^{+}$ effects with different values of $R_{Z\gamma}$ and the ratio of
$A_{0}(\eta,\kappa)/A_{0}(\tau)$ for the fixed $R_{\gamma\gamma} = 1.5$ are shown in Fig.~\ref{Fig_s2}. Here $R_{Z\gamma} = 0.5, 1, 2$ and $A_{0}(\eta,\kappa)/A_{0}(\tau)$
is taken to be $1.07$ and $1.00$ respectively. It shows that taking the ratio of $A_{0}(\eta,\kappa)/A_{0}(\tau)$ as a
constant is a good approximation in the regions of parameters we are interested in and is useful for the deviation
of $\Gamma_{Z\gamma}$ from the SM prediction for the singly charged singlet scalar $s^{+}$. Finally, it is worth to note that the charged singlet $s_1^+$ and $s_2^+$ can be produced by the quark annihilation through gauge boson $\gamma$ and $Z^0$ mediating. This process could have the final states including two charged leptons, the same as that of $pp\rightarrow h\rightarrow WW^*$, for which final states of $e$ or $\mu$ are tagged by LHC. The related signal strength observed from ATLAS is $1.3\pm0.5$~\cite{ATLAS-conf-158}, and the $1\sigma$ deviation of it corresponds to around $20\mathrm{pb}$. This deviation can be regards as the allowed space for new contribution beyond SM. The singly charged scalar production cross section with parton distribution function given from CTEQ6~\cite{Pumplin:2002vw} is also around $20\mathrm{pb}$ with $m_{s}\simeq 150\mathrm{GeV}$, which can be regarded as the lower bounds on $m_s$ if we assume $\mathrm{Br}(s\rightarrow e(\mu) \nu)\simeq 100\%$. Finally, we estimate
the background contribution for $h \rightarrow WW^{*} \rightarrow e\nu\mu\nu$ channel from $s^{\pm}s^{\mp}$ production. Both
singly charged scalars are off-shell and we found that the partial decay width for the center of mass in one of virtual scalar to
be around $m_{W}$ is negligible compared to the SM prediction.
%\begin{eqnarray}
%\,,\,\nonumber\\
%A_{1}(\tau)&=&-[2\tau^2+3\tau+3(2\tau-1)f(\tau)]\tau^{-2}\,,\,\nonumber\\
%A_{0}(\tau)&=&-[\tau-f(\tau)]\tau^{-2}\,.
%\end{eqnarray}

%\begin{eqnarray}
%A_{1/2}(\tau,\lambda)&=&I_1(\tau,\lambda)-I_2(\tau,\lambda)\,,\nonumber\\
%A_1(\tau,\lambda)&=&c_W\bigg\{4(3-{s_W^2\over c_W^2})I_2(\tau,\lambda)+[(1+{2\over\tau}){s_W^2\over c_W^2}-(5+{2\over \tau})]I_1(\tau,\lambda)\bigg\}\,,\nonumber\\
%A_0(\tau,\lambda)&=&I_1(\tau,\lambda)\,.
%\end{eqnarray}

%{\bf Q: Invisible axion mass from potential?}\\

%{\bf Q: express F by antisymmetric form of F'?}

\section{Discussions and Conclusions}
\label{sec:5}
We investigate a model that neutrinos are Dirac fermions and their masses are generated from the Peccei-Quinn symmetry
breaking at one-loop level. As a result the neutrino mass is related to $\Lambda_{\rm QCD}$ and axion is appeared to
be a good candidate of dark matter in explaining the missing energy density of our universe. Leptonic rare decays constrain
the model parameters severely, and therefore, the model can be tested in the near future. We also studied the implications of the
new scalars to the electromagnetic moments of leptons and recent Higgs signals at LHC, specifically the
$h \rightarrow \gamma\gamma$ and $h \rightarrow Z\gamma$ decays. Finally we would like to make a brief comment on the
Majorana extension of our scenario. Lepton number symmetry is one of the key to understand the underlying neutrino physics.
The smoking gun signals to resolve the question are $0\nu\beta\beta$ decays and some lepton number violating processes
at LHC. Without the direct observations Majorana neutrinos and Dirac neutrinos
are equally good in many aspects to describe the phenomena such as neutrino oscillations, leptogenesis, and Big Bang
Nucleosynthesis, etc. Although we discuss the Dirac neutrino in this model and argue that the Dirac neutrino mass
is originated from the Peccei-Quinn symmetry breaking which is well-motivated in QCD field theory, a Majorana masses
of the right-handed neutrinos, in general, can be formed without violating any principle except the lepton number symmetry. The
main modifications are to include some terms in the scalar potential such as $H_1H_2s_1^+$ and $H_1H_2as_2^+$.
Therefore, our discussions can be easily embedded the scenario of Majorana neutrino and diagonalize the neutrino mass matrix via
the seesaw mechanism.
\appendix
\section{Scalar Mass Spectrum}
We briefly analyze the scalar potential given in Eq.~(\ref{potential}) and their particle spectrum in this appendix. If the neutral components of $H_1=(h_1^+,\;(R_1+iA_1)/\sqrt{2})^T$, $H_2=(h_2^+,\;(R_2+iA_2)/\sqrt{2})^T$ and $a=(R_a+iA_a)/\sqrt{2}$ acquire the vacuum expectation values (VEVs) $v_1$, $v_2$, and $v_a$, respectively, and the related tadpole conditions can be expressed as 
\begin{eqnarray}
-\mu_1^2&=&-[\lambda_1v_1^2+{1\over2}(\lambda_3+\lambda_4)v_2^2+
{1\over2}(d_a+h_5{v_2\over v_1})v_a^2]\;,\nonumber\\
-\mu_2^2&=&-[\lambda_2v_2^2+{1\over2}(\lambda_3+\lambda_4)v_1^2+
{1\over2}(g_a+h_5{v_1\over v_2})v_a^2]\;,\\
-\mu_a^2&=&-[h_av_a^2+h_5v_1v_2+{1\over2}(d_av_1^2+g_av_2^2)]\;.
\end{eqnarray}
$v_{1,2}$ are responsible for the electroweak symmetry breaking and $v_a\approx 10^{12}$GeV is aimed for the PQ symmetry breaking scale. 
Notice that the sizes of $\mu_1$, $\mu_2$ and $\mu_a$ are not required to be in electroweak scale, while the TeV-scale charged singlets $s_{1,2}$ will bound the couplings $h_{a1,a2}$ to be around $v^2/v_a^2$ and the value of $\mu$ to be less than $(10^3\mathrm{GeV})^2/v_a$. Furthermore, the trilinear couplings $d_a$ and $g_a$ for the scalar fields and the axion will be constrained to be larger than $v/v_a$.
The $3\times3 $ scalar neutral mass matrix elements $M_{ij}$ in the basis $\{R_1,\;R_2,\; R_a\}$ are given as
\begin{eqnarray}
M_{11}^2&=&2\lambda_1(\cos\beta)^2 v^2-{h_5\over2}\tan\beta v_a^2\; , \\
M_{22}^2&=&2\lambda_2(\sin\beta)^2 v^2-{h_5\over2}\cot\beta v_a^2\;,\\
M_{33}^2&=&2h_av_a^2\;, \\
M_{12}^2&=&{1\over2}[(\lambda_3+\lambda_4)(\sin2\beta) v^2+h_5v_a^2]\;,\\
M_{23}^2&=&(g_a\sin\beta+h_5\cos\beta)v v_a\;, \\
M_{13}^2&=&(f_a\cos\beta+h_5\sin\beta)v v_a\,.
\end{eqnarray}
The relations $\tan\beta=v_2/v_1$ and $v=(v_1^2+v_2^2)^{1/2}$ are used in the formulae. The mixing angle between $R_{1,2}$ and mass eigenstates $H,H_h$ is defined as $\tan(\beta-\alpha)$, the same as the definition in \cite{Chang:2012ta}.
The discovery of $126$ GeV scalar implies that the magnitude of $h_5$ should be of the order $v^2/v_a^2$. Similarly, the masses of pseudo-scalar and charged Higgs bosons are given by
\begin{eqnarray}
m_A^2&=&-h_5(\sin2\beta v^2+{1\over \sin2\beta}v_a^2)\;,\nonumber\\
m_{H^\pm}^2&=&-{\lambda_4\over2}v^2-h_5{1\over \sin2\beta}v_a^2\;.
\end{eqnarray}
From the above mass formulae we get $h_5<0$, and $m_A^2-m_{H^\pm}^2=(\lambda_4/2)v^2$ by applying $v_a\gg v$. Finally, the charged particles $s_1$ and $s_2$ will not mix with $H^\pm$. The mass matrix elements $M_{sij}$ are given by
\begin{eqnarray}
M_{s11}^2&=&\mu_{s1}^2+h_{a1}v_a^2+{1\over2}[f_1(\cos\beta)^2+g_1(\sin\beta)^2]v^2\;,
\\
M_{s22}^2&=&\mu_{s2}^2+h_{a2}v_a^2+{1\over2}[f_2(\cos\beta)^2+g_2(\sin\beta)^2]v^2\;,
\\
M_{s12}^2&=&{\mu v_a\over\sqrt{2}}\;.
\end{eqnarray}
Notice that $h_{a1,a2}$ are chosen in the order of $(v/v_a)^2$ to ensure the electroweak scale $m_{s1}$ and $m_{s2}$. 
For simplicity if we take the mixing terms $M_{12}^2$, $M_{23}^2$, and $M_{13}^2$ to be vanished, we obtain 
\begin{eqnarray}
\lambda_3=-\lambda_4-{h_5v_a^2\over\sin2\beta v^2} \quad {\rm and} \quad
g_a=d_a=0\;.
\end{eqnarray}
Then $\lambda_3$ is also fixed once the values $\tan\beta$, $h_5$, and $\lambda_4$ are given, and at the moment $\sin{\alpha}=\sin\beta$. The trilinear couplings of SM Higgs $H$ with charged scalars $H^+$, $s_1$, and $s_2$ respectively are given by
\begin{eqnarray}
\mu_{HH^+H^-}&=&\sin\beta[2\lambda_2(\cos\beta)^2+
\lambda_3(\sin\beta)^2-\lambda_4(\cos\beta)^2] v\;,\;\nonumber\\
\mu_{Hs_1^+s_1^-}&=&g_1v\sin\beta \;,\;
\mu_{Hs_2^+s_2^-}=g_2v\sin\beta \;.\;
\end{eqnarray}

As an illustration, taking $\tan\beta=10$, $\lambda_1=0.5$, $\lambda_2=0.125$, $\lambda_4=0$, $\lambda_5=-0.3$ as input values, then the related scalar masses are $m_h=126~$GeV, $m_H\simeq m_A=m_{H^\pm}=303$GeV. On the other hand, for $\tan\beta=50$, $\lambda_1=\lambda_2=0.13$, $\lambda_4=0$, $\lambda_5=-0.1$, we have $m_H\simeq m_A=m_{H^\pm}=389$~GeV. Both cases satisfy the experimental constraints~\cite{Eberhardt:2013uba}. In summary, having the scalar with $126$~GeV mass is insensitive to the value of $h_5$ due to the $\cos\beta$ suppression. The current constraint on neutral scalar mass would give $|h_5(v_a^2/v^2)|\gtrsim 0.3$ for $\tan\beta=10$ and $|h_5(v_a^2/v^2)|\gtrsim 0.06$ for $\tan\beta=50$, by setting $\lambda_4=0$.

\acknowledgments
C-S~Chen is supported by National Center for Theoretical Sciences, Taiwan, R.~O.~C. and
L-H Tsai is supported by the National Tsing-Hua University (Grant No. 101N1087E1), Taiwan, R.O.C.

\end{document}